\documentclass{aastex}
\usepackage{emulateapj5,apjfonts,psfig}

\received{}
\accepted{}

\slugcomment{To appear in {\it The Astrophysical Journal Letters}}

\shortauthors{Halpern et al.}
\shorttitle{Multiwavelength Observations of 3EG J1835+5918}

\begin{document}

%
%
\def\xray{RX~J1836.2+5925}
\def\source{3EG J1835+5918}
\def\ro{{\it ROSAT\/}}
\def\asca{{\it ASCA\/}}

\title{The Next Geminga: Deep Multiwavelength Observations of
a Neutron Star Identified with 3EG J1835+5918}

\author{J. P. Halpern, E. V. Gotthelf, N. Mirabal, and F. Camilo}
\affil{Columbia Astrophysics Laboratory, Columbia University, 550 West 120th Street,
 New York, NY 10027}

\begin{abstract}
\rightskip 0pt \pretolerance=100 \noindent
We describe {\it Chandra}, {\it HST\/},
and radio observations that reveal a radio-quiet but magnetospherically
active neutron star
in the error circle of the high-energy $\gamma$-ray source \source,
the brightest of the unidentified EGRET sources
at high Galactic latitude.
A {\it Chandra} ACIS-S spectrum of the ultrasoft X-ray source
\xray, suggested by Mirabal \& Halpern as the
neutron star counterpart of \source,
requires two components: a blackbody of $T_{\infty} \approx 3 \times 10^5$~K
and a hard tail that can be parameterized as a power law of photon
index $\Gamma \approx 2$.  An upper limit of $d < 800$~pc
can be derived from the blackbody fit under an assumption of
$R_{\infty} = 10$~km.  Deep optical imaging with the {\it HST\/}
STIS CCD failed to detect this source to a limit of $V > 28.5$,
thus $f_X/f_V > 6000$ and $d > 250$~pc assuming the X-ray fitted
temperature for the full surface.  Repeated
observations with the 76~m Lovell telescope at Jodrell Bank
place an upper limit of $< 0.1$~mJy on the flux density 
at 1400~MHz for a pulsar with $P>0.1$~s, and $<0.25$~mJy for
a $\sim 10$~ms pulsar
at the location of \xray.  All of this evidence points to
an older, possibly more distant version of the highly efficient
$\gamma$-ray pulsar Geminga, as the origin of
the $\gamma$-rays from \source.

\end{abstract}


\section{Introduction}

The nature of the persistent high-energy ($> 100$ MeV)
$\gamma$-ray sources in the Galaxy remains a puzzle
three decades after their discovery.  Ever since the
identification of the mysterious $\gamma$-ray source Geminga as the
first radio quiet but otherwise ordinary pulsar (see review by
Bignami \& Caraveo 1996), it has been argued
that rotation-powered pulsars likely dominate
the Galactic $\gamma$-ray source population
(e.g., Halpern \& Ruderman 1993; Helfand 1994;
Kaaret \& Cottam 1996; Yadigaroglu \& Romani 1997),
and that many of them are radio quiet.
That Geminga is a bright EGRET source at a distance of only $\sim 200$~pc
(Halpern \& Ruderman 1993; Caraveo et al. 1996) begs for it
not to be unique.

As a $\gamma$-ray pulsar spins down, an increasing fraction
of its power is emitted above 100~MeV (Thompson et al. 1997, 1999).
This implies that the $\gamma$-ray source population is not
necessarily dominated by {\it young} pulsars.
Like Geminga and PSR~B1055--52,
many of them could be quite old, up to $10^6$~yr or more.
Geminga itself has a characteristic age of $3.4 \times 10^5$~yr,
and the EGRET pulsar PSR~B1055--52 is $\approx 5 \times 10^5$~yr old.
The unidentified $\gamma$-ray sources may therefore include
a significant fraction of the nearby neutron-star population
that has yet to be accounted for, and may provide a useful
window into their physics.  It is also speculated
(Gehrels et al. 2000; Grenier 2000; Harding \& Zhang 2001)
that as many as 40 of the steady, unidentified EGRET sources at
intermediate Galactic latitude are a population of older pulsars
born in the Gould Belt, an inclined, expanding disk of star formation
in the solar neighborhood that is $\approx 3 \times 10^7$~yr old.
If so, they will most easily be identified as
soft blackbody X-ray sources without optical counterparts.

As the brightest of the as-yet unidentified high-latitude EGRET sources
(Hartman et al. 1999), 
at $(\ell, b)=(89^{\circ},+25^{\circ})$, \source\ is a good a priori
target for the next such identification, and a candidate for the prototype
of the putative Gould Belt pulsars.
Mirabal et al. (2000, hereafter Paper I)
performed an exhaustive, multiwavelength search for a counterpart of \source. 
They identified optically all but one of the {\it ROSAT\/}
and {\it ASCA\/} sources in the region of \source\
to a flux limit of
$\sim 5 \times 10^{-14}$~ergs~cm$^{-2}$~s$^{-1}$, which is $10^{-4}$
of the $\gamma$-ray flux, without finding a
plausible counterpart among the identified
sources.  Mirabal \& Halpern (2001, hereafter Paper II) concluded
that the one optically undetected X-ray source, \xray\ (also the brightest
one in the EGRET error circle), is the most promising
candidate for identification with \source\ principally because
of the absence of optical emission.
The upper limit (at that time) on the optical flux from \xray\ implied
that its ratio of X-ray-to-optical flux $f_X/f_V$ is greater than 300,
an extreme that is seen only among neutron stars. 
The detection of \xray\ as
a weak, ultrasoft source in the \ro\ All-Sky Survey (Paper II), 
further suggested that it is a thermally emitting neutron star that is either
older or more distant than Geminga.

We designed follow-up observations with
the {\it Chandra X-ray Observatory} and the {\it Hubble Space Telescope}
({\it HST\/}) to test this hypothesis and to further characterize the
properties of the presumed neutron star.  We
also re-examined its location for evidence of a radio pulsar.
Previous searches of the \source\ error box failed to find a radio
pulsar to a limit of 1~mJy at 770~MHz (Nice \& Sayer 1997),
and radio continuum observations of
the entire error box using the VLA at 1420~MHz yielded an upper limit
of 0.5~mJy at the location of \xray\ (Paper I).  These limits are not below
the range of several radio pulsar detections, so we performed deeper
searches at Jodrell Bank to improve upon them.

\section{Chandra X-ray Observation}

A {\it Chandra} observation with the Advanced CCD Imaging Spectrometer
(ACIS; Burke et al. 1997) was made on 2002 March 6.  The target source
\xray\ was positioned at the default position on
the back-illuminated S3 chip of the ACIS-S array.
The total usable exposure time was 28,117~s.
Figure~1 shows a smoothed version of the image.
A total of 260 counts were extracted from a $1\farcs9$
radius circle centered on \xray\ using the CIAO
``psextract'' script.  A spectral file was produced by grouping
at least 20 counts per channel.  Instrument and mirror response 
files were generated using the CIAO tools ``makermf'' and ``makearf''.
The expected contribution of background to the source spectrum is
only $\approx 4$ counts, so no background subtraction was performed.
There is no evidence for extended emission (synchrotron nebulosity)
associated with this source.

Figure~2 shows the grouped ACIS spectrum, which is clearly that
of an ultrasoft source, as first revealed by the {\it ROSAT\/}
All-Sky Survey (Paper II).  The fact that the counts are
rising down to the lowest energy signifies both a soft intrinsic
spectrum and little absorbing column.
Although the ACIS response below 0.6 keV may not be well calibrated
(e.g., Zavlin et al. 2002), we do not have the freedom to exclude these
energies from spectral fitting because they contain the majority of
our photons.  Instead, we proceed
with the following analysis based on the current calibration.
First, we ignore the spectral channel
around 0.28~keV, since a deviant point there may be associated with
poor instrumental calibration around the carbon K-edge.  Next,
attempts to fit the spectrum using a blackbody or power-law model alone
produced unsatisfactory fits, with $\chi^2$/dof = 10 and 4, respectively.
This result is insensitive to $N_{\rm H}$, whether the latter is
treated as a free parameter or held fixed at the maximum Galactic
value in this direction ($4.6 \times 10^{20}$~cm$^{-2}$). A two-component
model consisting of a blackbody plus a power law produced adequate
fits, with reduced $\chi^2 < 1.4$ for all values of
$N_{\rm H} \leq 4.6 \times 10^{20}$~cm$^{-2}$.  However, the small number of
photons, extremely soft spectrum, and unknown intervening column density
render the blackbody flux and bolometric luminosity highly uncertain
even if the instrument response were known accurately.
Rather than leaving $N_{\rm H}$ a free parameter in the fits, we explore
the effects of assuming particular values
of $N_{\rm H}$ ranging from $2.5 \times 10^{19}$~cm$^{-2}$
to the maximum Galactic value of $4.6 \times 10^{20}$~cm$^{-2}$.
Figure~3 shows the corresponding confidence contours.  The
best fitted blackbody temperatures for this range of $N_{\rm H}$ vary
from $(2.9 - 3.5) \times 10^5$~K, with a $1\sigma$ upper limit of
$5.5 \times 10^5$~K.
The high-energy tail that is required to accompany the blackbody
component is parameterized as a power-law of photon index
$1.6 \leq \Gamma \leq 2.8$ ($1\sigma$ limits), with flux in the $0.2-2.0$~keV
band of $(1.9 - 2.6) \times 10^{-14}$ ergs~cm$^{-2}$~s$^{-1}$.

We bound the range of plausible distances
by assuming a blackbody radius of 10~km for each of the trial
values of $N_{\rm H}$.  Under this assumption, $100 \leq d \leq 800$~pc,
and $5 \times 10^{30} \leq L_{\rm Bol} \leq 1.1 \times 10^{31}$
ergs~s$^{-1}$; smaller distance corresponds to
smaller temperature.  Formally, no
lower limit on $T_{\infty}$ can be set from the X-ray spectrum alone,
although the distance would become unreasonably small for
$T_{\infty} < 3 \times 10^5$~K
given the lack of optical detection, which sets an additional
lower bound on $d$ (see \S 3).

Precise astrometry of the ACIS image
in the USNO-A2.0 optical
reference frame was achieved by
registration of four previously identified, ``bright'' X-ray sources in
Figure~1 with their optical counterparts on the MDM 2.4m
$V$-band CCD image described in Papers I and II.  This required
a zero-point shift to the {\it Chandra} aspect solution
of $-0\farcs13$ in right ascension
and $-0\farcs39$ in declination,
which is within specifications.  The corrected
position of \xray\ is (J2000)
$18^{\rm h}36^{\rm m}13.\!^{\rm s}723,
+59^{\circ}25^{\prime}30\farcs05$,
which should be accurate to $0\farcs1$
relative to our ground-based optical image
as indicated by the residual dispersion in X-ray-optical
offsets of the identified sources.  We note that this
position is only $0\farcs5$
from the {\it ROSAT\/} HRI position that was determined
in a similar manner in Paper II, confirming that the
latter's error circle radius of $3^{\prime\prime}$
was conservative.  The agreement in position between
these two X-ray observations that were made 4 years
apart already
limits any proper motion of the source to $< 1^{\prime\prime}$ yr$^{-1}$.

\medskip
\centerline{
\psfig{file=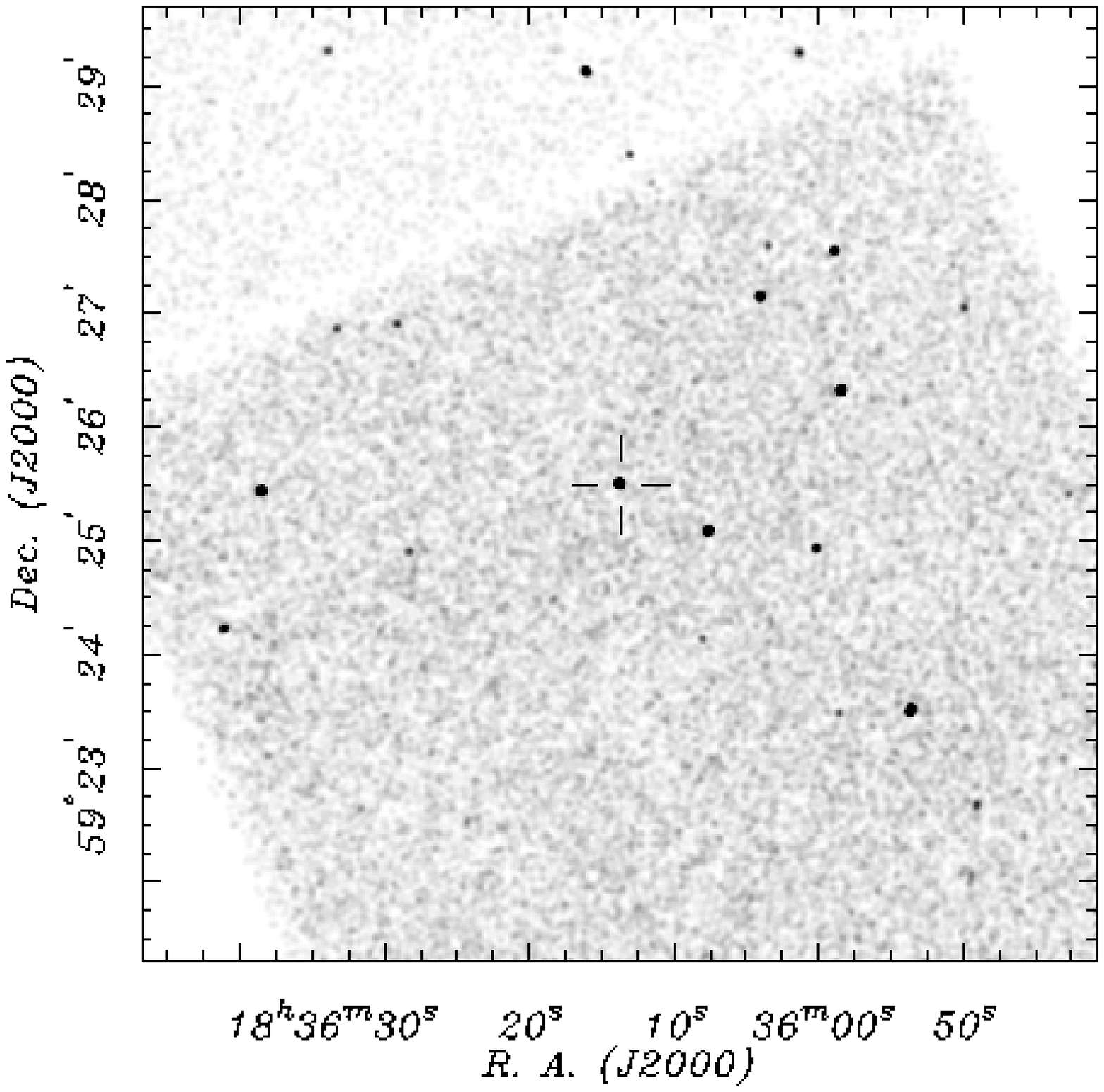,height=3.3in,angle=-90.}
}
{\footnotesize FIG. 1.--- Smoothed {\it Chandra} ACIS-S3
image centered on the source \xray\ ({\it cross}).
}

\centerline{
\psfig{file=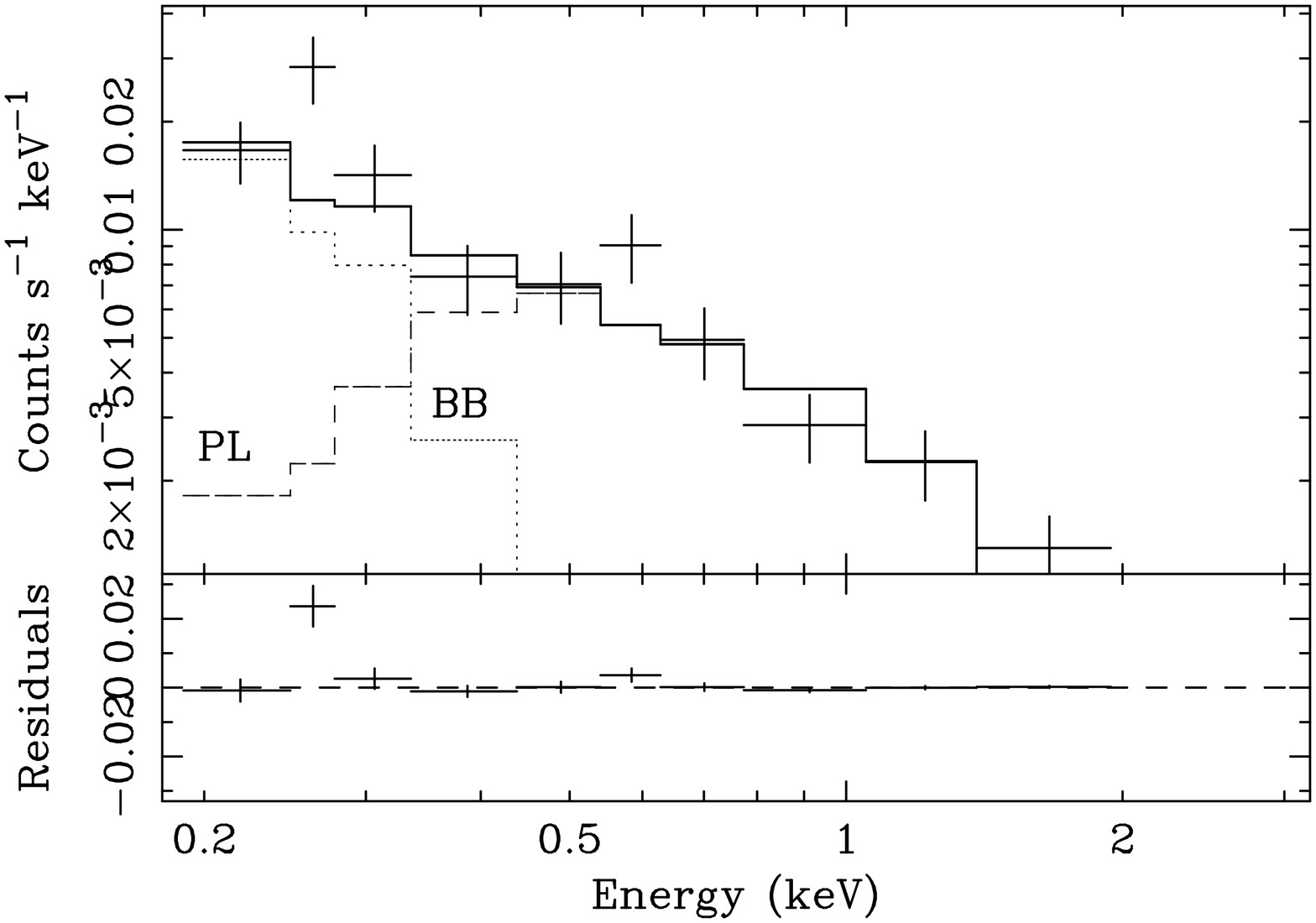,height=2.8in,angle=-90.}
}
{\footnotesize FIG. 2.--- {\it Chandra} ACIS-S3 spectrum of
\xray, the neutron star counterpart of \source.  {\it Top panel}:
Data ({\it crosses}) and best-fit model ({\it thick line})
for an assumed $N_{\rm H} = 4.6 \times 10^{20}$ cm$^{-2}$
(see text).  Contributions of the
blackbody ({\it dotted line}) and power law ({\it dashed line})
components are shown. {\it Bottom panel}: Difference between data and
model, in the same units as the top panel.
}

\section{HST STIS CCD Imaging}

Deep images were obtained on 2002 February 2 using the 
{\it Space Telescope Imaging Spectrograph\/} (STIS) CCD
on {\it HST\/},
both with open filter (50CCD) and long-pass filter (F28X50LP).
In each filter, a total exposure time of 10,400~s was obtained
in a standard parallelogram dither pattern with four exposures at

\centerline{
\psfig{file=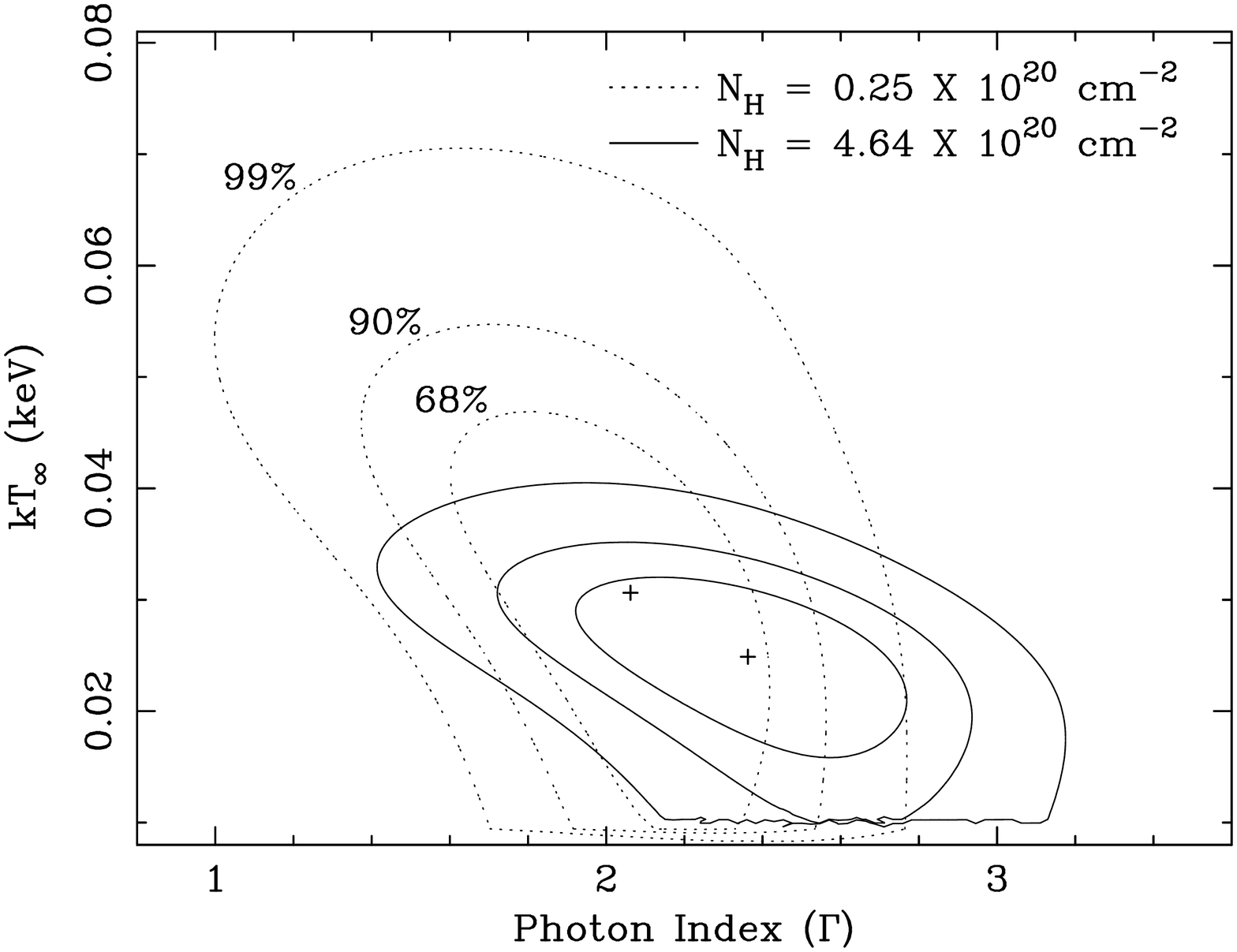,height=2.6in,angle=-90.}
}
{\footnotesize FIG. 3.---  Confidence contours for the
two-component fit ($kT_{\infty}, \Gamma$)
to the {\it Chandra} spectrum of
\xray.  The intervening column density has been
held fixed at two different values,
the maximum Galactic value being $N_{\rm H}
= 4.6 \times 10^{20}$ cm$^{-2}$.  Confidence levels
are for two interesting parameters.
}
\medskip

\noindent
each point to facilitate the rejection of cosmic rays.
The data were processed using the standard STIS pipeline,
registered, and combined according to the dither pattern.
Figure~4 shows a portion of the 
resulting open filter image centered on \xray.
Aperture photometry was used to obtain STIS magnitudes from
the PHOTLAM and PHOTZPT keywords in the image headers.
We then adopted the transformation equations derived by
Rejkuba et al. (2000) from observations of stars in the 
irregular galaxy WLM to convert STIS magnitudes in the open
and long-pass filters to the $V$ and $I$ Kron-Cousins
system.  We also examined a set of objects in the STIS images
that we had previously calibrated in ground-based photometry,
and the zero-points agree.  Thus, we find that the limiting
magnitudes for point-source detection at the $3\sigma$ level
are $V = 28.8$ from the 50CCD image and and $I = 26.5$
from the F28X50LP.  The main uncertainty in
this calibration is due to the broad passbands of the STIS
filters, and is estimated as 0.2 mag.

In order to tie the {\it HST\/} and {\it Chandra} images to
the same astrometric reference frame, we used
the positions of 10 objects that are present on
the both the {\it HST\/} STIS and ground-based CCD images
to transfer the USNO-A2.0 reference frame to
the STIS image.  These objects include
stars and compact galaxies.  The dispersion
among these 10 secondary astrometric standards from the fit to their
{\it HST\/} positions is $0\farcs065$,
or slightly larger than 1 STIS pixel.  Thus, the combined
uncertainty in {\it Chandra}
and {\it HST\/} astrometry is less than $0\farcs2$;
we use a conservative error circle of radius
$0\farcs2$ in Figure~4
to indicate the location of the X-ray source on the STIS image.
Since the {\it HST\/} and {\it Chandra} observations
were made only one month apart, we are confident that
any proper motion, already limited to $< 1^{\prime\prime}$ yr$^{-1}$,
is of no importance.

The {\it Chandra} error circle excludes all of the optical objects within
the {\it ROSAT\/} error circle that were detected by Totani, Kawasaki,
\& Kawai (2002) in their $B$-band image obtained on the Subaru telescope.
At the northeast edge of the
{\it Chandra} error circle there is only a marginal source of
$V \approx 29.0 \pm 0.4$ in the STIS 50CCD image, but it is not
present in the F28X50LP image.
Since this is not even a $3\sigma$ detection, we consider
that the X-ray source is formally undetected optically,
with upper limits of $V > 28.5$ and $I > 26.5$,
and that it must therefore be a neutron star with
$f_X/f_V > 6000$.
The absence of an optical detection places additional constraints
on the distance and temperature of the neutron star \xray.
A magnitude limit of $V > 28.5$ corresponds to a flux
$< 0.014\,\mu$Jy at a wavelength of 5500 \AA, whereas the Rayleigh-Jeans
flux from a neutron star at that wavelength would be
$0.030\,T_5\,(R_{10}/d_{100})^2\,\mu$Jy,
where $T_5$ is $T_{\infty}$ in units
of $10^5$~K, $R_{10}$ is the radius
in units of 10~km, and $d_{100}$ is the distance in units of 100~pc.
Therefore, the nominal X-ray fitted temperature of $3 \times 10^5$~K,
if coming from the full surface of the neutron star, would require
$d > 250$~pc.

\medskip
\centerline{
\psfig{file=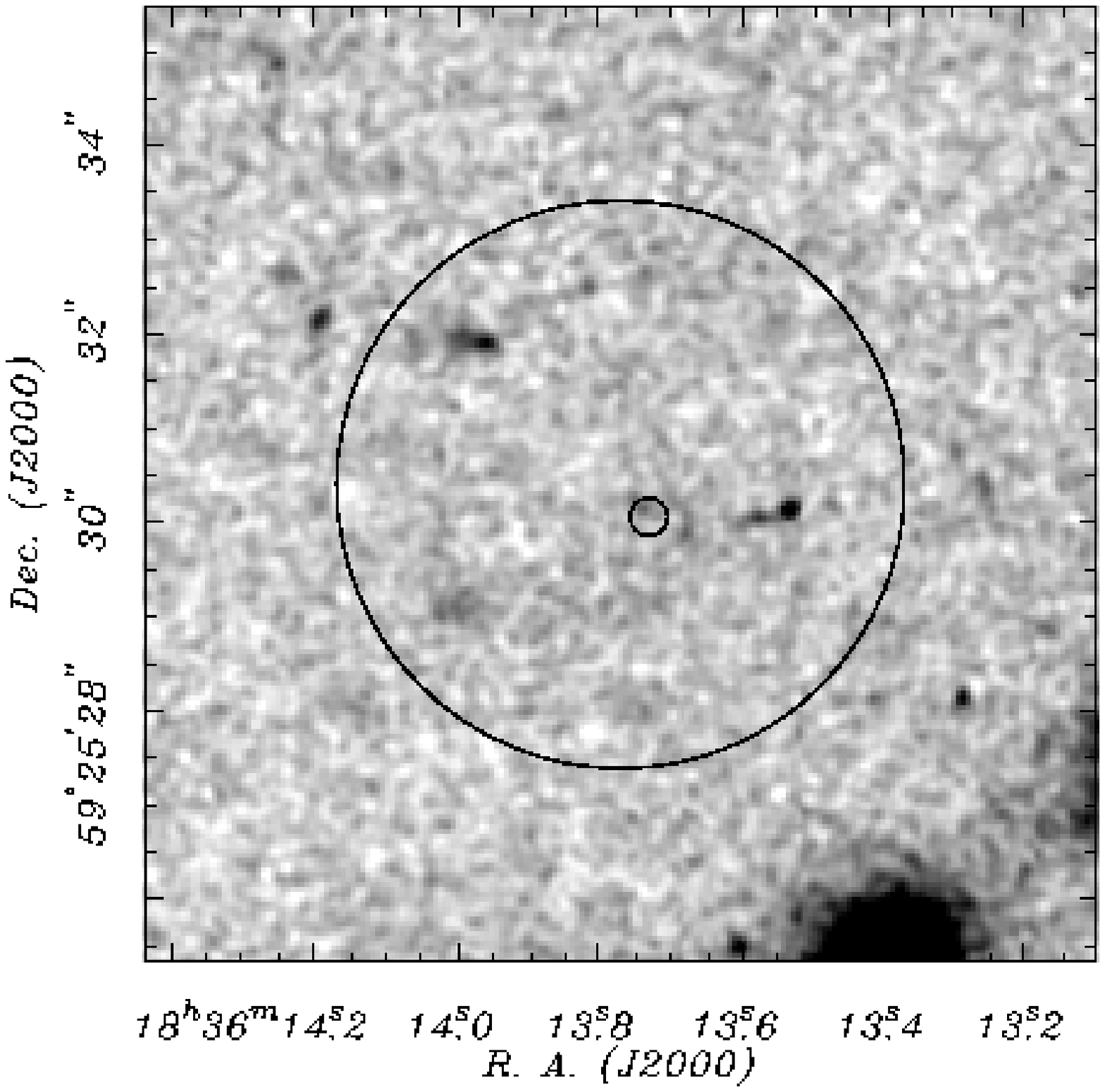,height=3.3in,angle=-90.}
}
\medskip
{\footnotesize FIG. 4.--- Smoothed, open filter {\it HST\/} STIS image
centered on the X-ray source \xray.  The $3^{\prime\prime}$ radius
error circle from the {\it ROSAT\/} HRI, and the $0\farcs2$
radius {\it Chandra} error circle are shown.  The limiting magnitude
of this image is equivalent to $V = 28.8$.
}
\medskip

\section{Radio Pulsar Search}

The absence of a ``strong,'' i.e, $\sim 0.5$ mJy radio source at the
position of \xray\ was already known from Paper I.  In order to
make a more sensitive search for radio pulsations, we obtained six
observations on different days in 2001 February and March for
2.3~hr each with the Jodrell Bank 76~m Lovell telescope at
a frequency of 1400~MHz.
A $64\times 1$\,MHz filter bank and a sampling time of 1~ms were used.
Since the maximum expected
dispersion measure is small at this high-latitude location, DM $\leq
17$ pc~cm$^{-3}$ for $d \leq 1$~kpc
according to the Taylor \& Cordes (1993) model,
the limiting flux density for pulsations is insensitive to distance
or period in the range $P > 100$~ms, being 0.1~mJy for a duty
cycle of 10\%.  A period in this range is expected for an older,
Geminga-like pulsar.
For $P < 100$~ms, the sensitivity diminishes
rapidly, to $\approx 0.25$~mJy at 10~ms.  Scintillation might
be expected for these DM values and observing parameters,
but our six observations should adequately sample most any such
variability.  Finally, we obtained a single
9.3~hr pointing on 2002 February 18 with the same sampling as before,
which reduces the nominal sensitivity to $50\,\mu$Jy, although
scintillation may play a role here.  No pulsations
were detected in any observation.

Adopting an upper limit of 800~pc to the distance from the
blackbody fit to the X-ray spectrum, a conservative upper
limit on the pulsed radio luminosity of \xray\ is $L_{1400}
< 0.064$~mJy~kpc$^2$.
The least luminous young pulsar known in radio is the one
in the supernova remnant
3C~58, with $L_{1400} \approx 0.5$~mJy~kpc$^2$ (Camilo et al. 2002a),
and there are only four pulsars known with $L_{1400} < 0.1$~mJy~kpc$^2$
(Camilo et al. 2002b)
in addition to Geminga (McLaughlin et al. 1999, and references
therein).  \xray\
is certainly in contention as a radio-quiet pulsar,
although it would not be surprising if more sensitive
observations discover it at slightly lower flux density
than explored here.

\section{Discussion and Conclusions}

The results of these deep X-ray, optical, and radio observations are
quite revealing of the properties of \xray.  Without exception
they support the hypothesis that it is an older and possibly more distant
cousin of the Geminga pulsar, and readily identifiable with the EGRET source
\source\ in which error circle it lies.  First, the X-ray source,
which is primarily an ultrasoft blackbody of $T_{\infty}
\approx 3 \times 10^5$~K,
is significantly cooler than the oldest ``ordinary''
$\gamma$-ray pulsars Geminga and PSR B1055--52,
which have
$T_{\infty} = (5.6 \pm 0.6) \times 10^5$~K and $(7.5 \pm 0.6) \times 10^5$~K,
respectively (Halpern \& Wang 1997; \"Ogelman \& Finley 1993).  Most of the
neutron star cooling curves that match the temperatures
of Geminga and PSR B1055--52 drop to $T_{\infty} = 3 \times 10^5$~K at
an age $\geq 1 \times 10^6$~yr (e.g., Yakovlev et al. 2002),
approximately 3 times the characteristic
age of Geminga.  At the same time, the difference in temperature
is not sufficient to account for the factor of 40 difference in
soft X-ray flux between Geminga and \xray, so the latter may
be as far away as 800~pc, as compared to $d \sim 200$ pc for Geminga.
At the maximum
distance, \xray\ would be 340~pc from the Galactic plane, not an
unreasonable distance for a neutron star to have traveled
in $10^6$~yr (Paper II).

Even more important is the presence of an apparently nonthermal
extension to the X-ray spectrum.  All of the EGRET pulsars, even
the oldest ones, have such ``power-law'' components that can be explained
in terms of synchrotron emission from secondary particles that
are produced in the conversion of primary $\gamma$-rays in the strong
magnetic field near the surface of the neutron star (Wang et al. 1998).
No $\gamma$-ray pulsar lacks such a nonthermal X-ray spectrum,
while several middle-aged or older pulsars that might be expected
to be EGRET sources based on their proximity and/or spin-down
power have only thermal X-ray spectra and no $\gamma$-ray emission.
That a neutron star as cool as \xray\ possesses a non-thermal
X-ray tail is encouraging of a connection with \source.

All of the non-thermal manifestations of
\source\ are fainter than the corresponding emission from
Geminga: by a factor of $\geq 20$ in the optical, $\approx 15$
in the X-ray, and $\approx 6$ in the $\gamma$-ray.  Its thermal
X-ray flux, although difficult to quantify,
is also much less than that of Geminga.  While \xray\
is relatively inconspicuous, it {\it is} the brightest soft
X-ray source within the error circle of \source.  If a fainter X-ray
source were considered as a possible counterpart,
it would only exacerbate the differences between
this and other, identified EGRET sources as
discussed in Paper I.

Finally, we mention once again the two pieces of evidence
that suggest \source\ itself is a priori
more likely to be a pulsar rather than, e.g., a blazar, the other
major class of EGRET source.  According to
Reimer et al. (2001) it shows no evidence for long-term
variability, and its spectrum can be fitted
by a relatively flat power law of $\Gamma = 1.7$ from 70~MeV to
4~GeV, with a turn-down above 4 GeV.  In contrast, blazars
are highly variable, and tend to have steeper spectra.
The energetics of \source\ are also plausible for a pulsar
at $d <  800$~pc since its $\gamma$-ray luminosity
(assumed isotropic) is $3.8 \times 10^{34}\,(d/800\,{\rm pc})^2$ ergs~s$^{-1}$,
comparable to the spin-down power $I\Omega \dot\Omega$ of Geminga
($3.3 \times 10^{34}$ ergs~s$^{-1}$).  Efficiencies approaching 100\%
are achieved by the least luminous $\gamma$-ray pulsars.
Fortunately, several future
observations have the potential to confirm this scenario in detail.
A pulsar detection is possible in a long observation with {\it XMM-Newton}
or with the {\it Chandra} High-Resolution Camera
if \xray\ is an ``ordinary'' pulsar, with $P > 50$~ms
and X-ray pulsed fraction $\geq 15\%$, as is common for thermally emitting
neutron stars that are also EGRET sources.  We have also not given
up on obtaining a radio pulsar detection.  Finally, the {\it Gamma-ray
Large Area Space Telescope} (GLAST) may itself be able to detect
pulsations from \source\ if a sufficiently long observation is made.

\acknowledgements

We thank Andrew Lyne for obtaining the radio data at Jodrell Bank,
and the referee Patrizia Caraveo for helpful suggestions.
This work was supported by grants SAO GO2-3071X and
HST GO-09278.01A.  The ability to obtain coordinated
{\it Chandra} and {\it HST\/} observations under a joint
proposal opportunity enabled us
to achieve these results in a timely and efficient manner.

\end{document}